\documentclass[12pt]{article}

\usepackage{amsmath,amsfonts,amssymb,amsthm,latexsym}
\usepackage[textures]{graphics}
\usepackage{epsf}

\def\be{\begin{equation}}
\def\ee{\end{equation}}

\def\beqa{\begin{eqnarray}}
\def\beqas{\begin{eqnarray*}}
\def\eeqa{\end{eqnarray}}
\def\eeqas{\end{eqnarray*}}
\def\bea{\begin{eqnarray}}
\def\eea{\end{eqnarray}}
\def\nn{\nonumber \\}
\def\xmy{x - y}
\def\br{\begin{eqnarray}}
\def\er{\end{eqnarray}}

\def\eps{{\epsilon}}
\def\ie{{\it i.e.\/}}

\def\a{\alpha}
\def\b{\beta}
\def\al{\alpha}

\def\th{\theta}

\def\d{\delta}

\def\med{{\frac{1}{2}}}

\def\bnabla{\boldsymbol{\nabla}}
\def\bbnabla{\bar{\boldsymbol{\nabla}}}
\def\bGamma{\boldsymbol{\Gamma}}

\def\bOmega{{\boldsymbol{\Omega}}}
\def\bbOmega{\bar{\bOmega}}
\def\bW{{\bf W}}
\def\bbW{\bar{\bf W}}

\def\Tr{{\rm Tr}}

\def\d{\partial}

\def\cpc{4\pi^2}

\def\sqr#1#2{{\vcenter{\vbox{\hrule height.#2pt
         \hbox{\vrule width.#2pt height#1pt \kern#1pt
            \vrule width.#2pt}
         \hrule height.#2pt}}}}
\def\dal{\mathop{\mathchoice\sqr64\sqr64\sqr{3.75}4\sqr34}\nolimits}

\makeatletter
\def\secteqno{\@addtoreset{equation}{section}%
\def\theequation{\thesection.\arabic{equation}}}

\begin{document}
\secteqno

\renewcommand{\thefootnote}{\fnsymbol{footnote}}

\noindent

\begin{titlepage}
\begin{flushright}
{ ~}\vskip -1in
 DFPD 02/TH/03\\
{\tt hep-th/0202082 }\\
February 2002\\
\end{flushright}

\vspace*{20pt}
\bigskip

\begin{center}
{ \Large \bf The beta function of N=1 SYM \\ in Differential
Renormalization }
\end{center}

\bigskip

\vskip 0.8truecm

\centerline{\sc
Javier Mas$^{\,a,}$\footnote{\tt jamas@fpaxp1.usc.es},
Manuel P\'erez-Victoria$^{\,b,}$\footnote{{\tt manolo@pd.infn.it}. On
leave of absence from
Depto.\ de F\'{\i}sica Te\'orica y del Cosmos, Universidad de
Granada, 18071 Granada, Spain. }
and Cesar Seijas$^{\,a,}$\footnote{\tt cesar@fpaxp2.usc.es}}

\vspace{1pc}

\begin{center}
{\em

 $^a$ Departamento de F\'\i sica de Part\'\i culas, Universidade de
Santiago de Compostela,\\
E-15706 Santiago de Compostela, Spain\\

\bigskip
$^b$ Dipartimento di Fisica ``G. Galilei'', Universit\`a di Padova
and \\
INFN, Sezione di Padova, Via Marzolo 8, I-35131 Padua, Italy} \\

\vspace{5pc}

{\large \bf Abstract}

\end{center}

\noindent
Using differential renormalization, we calculate the complete
two-point function of the background gauge superfield in pure N=1
supersymmetric Yang-Mills theory to two loops. Ultraviolet and
(off-shell) infrared divergences are renormalized in position  and
momentum space respectively. This allows us to reobtain the beta
function from the dependence on the
ultraviolet renormalization scale in an infrared-safe way. The
two-loop coefficient of the beta function is generated by the
one-loop ultraviolet renormalization of the quantum gauge field via
nonlocal terms which are infrared divergent on shell. We also discuss
the connection of the beta function to the flow of the Wilsonian
coupling.

\end{titlepage}

\renewcommand{\thefootnote}{\arabic{footnote}}
\setcounter{footnote}{0}


\section{Introduction}

In this paper we revisit a somewhat old controversy: the origin
of higher-order perturbative contributions to the beta function in
supersymmetric gauge theories. The relevance of infrared (IR) modes
and the distinction between the Wilsonian action and the generator of
1PI functions are at the heart of this debate. A good understanding of
these issues is relevant, for instance, for the comparison of
field theory results with holographic renormalization group
flows.\footnote{The calculation of the beta function of N=1 SYM in the
context of the AdS/CFT correspondence has been recently considered
in~\cite{Herzog:2001xk}. In order to discriminate between the different
possible results (which depend on the detailed UV/IR relation one
uses), a preliminar requisite is to identify the field
theory answer one would like to reproduce. This requires
a correct field-theoretical interpretation of the r\^ole
of the IR degrees of freedom.}

The so-called ``exact beta function'' of general N=1 SYM was
discovered by Novikov, Shifman, Vainshtein and Zakharov (NSVZ) using
instanton analysis~\cite{Novikov:uc}. In  this approach it is clear that
corrections to the one-loop result have an IR
origin in an imbalance in the number of fermionic and
bosonic zero modes. For pure SYM the NSVZ beta function reads
\begin{equation}
\beta(g)_{_{NSVZ}} = \frac{-3 C_A}{16\pi^2} \frac{g^3}{1-\frac{C_A
g^2}{8\pi^2}} \label{NSVZ} \; .
\end{equation}
This formula was first derived by Jones in~\cite{Jones:ip}.
Later on, superspace perturbative computations to two
loops were carried out using regularization by dimensional reduction
(DRed)~\cite{Abbott:pz,Grisaru:ja,Grisaru:1985tc}. In
this method the two-loop correction to the beta function arises from a
local evanescent operator specific to DRed\footnote{This is the
only local gauge invariant operator of the appropriate dimension: ${\rm
Tr} \int d^4x  d^4 \theta \Gamma^a \Gamma_b
(\delta_a^b-\hat{\delta}_a^b)$, where
$\delta$ and $\hat{\delta}$ are Kronecker deltas in four and $n$
dimensions, respectively. Using a Bianchi identity it can be cast in a
form proportional to the classical action: $-\eps {\rm Tr} \int d^4x
d^2\theta W^aW_a$.}. This operator is not available in regularization
schemes that stay in four dimensions and Grisaru, Milewski and Zanon
pointed out that this seems to imply that no
divergence should occur beyond one loop, in conflict with the DRed
result~\cite{Grisaru:1985tc}. NSVZ then observed that in four
dimensions the higher-loop
corrections can only arise from nonlocal operators that are
nonanalytic at vanishing external momentum~\cite{Novikov:rd}. This
behaviour can appear
only from the domain of virtual momenta of order of the external
momenta. Since this domain is excluded by definition in the Wilson
action, the flow of the Wilsonian coupling constant is purely
one-loop~\cite{Shifman:1986zi}.  The
standard modern proof of this fact is based on the holomorphic
dependence on the
complexified coupling constant. The running of the
physical coupling constant, on the other hand, has higher-order
contributions that appear when one takes the expectation value (in an
external field) of the operators in the Wilson action. Furthermore,
the IR pole is related to an anomaly, and this is the crucial fact
that allows to determine exactly the higher order
contributions~\cite{Shifman:1986zi}. In a later contribution,
Arkani-Hamed and Murayama rederived the NSVZ
beta function in
a purely Wilsonian setup~\cite{Arkani-Hamed:1997mj}. They showed
that the {\em canonical} Wilsonian coupling constant obeys a NSVZ flow.
The reason is that keeping canonical  kinetic terms at each scale requires
a rescaling of gauge  field which is anomalous. This
anomaly generates the higher order corrections. These authors claimed that
their calculation only depends on ultraviolet (UV) properties of the
theory, and thus questioned the IR origin of the corrections. Furthermore,
they pointed out that the method of differential
renormalization~\cite{Freedman:1991tk}
clearly displays an UV origin of the corrections. This
interpretation has been criticized in~\cite{Shifman:1999kf}.

The purpose of this paper is to try to contribute to the understanding
of these issues by performing an explicit calculation in differential
renormalization (DiffR)~\cite{Freedman:1991tk}. The interest of using
this method is twofold: on the one hand we are able to derive the beta
function directly from the scale dependence of finite renormalized Green
functions, rather than from ``infinite'' counterterms; in this
derivation we see explicitly how nonlocal terms contribute to the beta
function in a perturbative four-dimensional method. On the
other hand, we clearly separate UV divergences from the off-shell IR
divergences that afflict these calculations\footnote{In the
dimensional methods UV and IR
divergences are mixed. In~\cite{Abbott:pz} the IR divergences were
removed by the choice of a nonlocal gauge fixing that kept the
renormalized quantum propagator in the Feynman
gauge. In~\cite{Grisaru:1985tc} the IR divergences
were simply subtracted by an $\tilde{R}$
operation~\cite{Chetyrkin:nn}. We
shall also perfom an $\tilde{R}$ operation, but keeping track of the
resulting finite part.}.
As a by-product, we develop some calculational tools that
we believe have an intrinsic interest to the SUSY-community.
In fact, DiffR is a computational program that
seems to be especially taylored for supersymmetric theories:
it neither requires continuation in spacetime dimensions, nor
changing the field content of the bare lagrangian. It is rather
an implementation of Bogoliubov's $R$
operation, which yields directly renormalized correlation functions
satisfying renormalization group equations.
DiffR has been applied before to supersymmetric
calculations in the Wess-Zumino model~\cite{Haagensen:1991vd} to three
loops, in
pure SYM and SQED to one and two loops, respectively~\cite{Song}, and
in supergravity to one loop~\cite{delAguila:1997yd}. Its
implementation in symbolic programs~\cite{Hahn:1998yk} enables
efficient one-loop calculations in more involved models like the
MSSM~\cite{Hahn:2001rv}.

The layout of the paper is the following. In Section~\ref{sec_DR} we
review the method of DiffR and introduce the new tools needed
for our calculation. In Section~\ref{sec_covariant} we quickly review
the supercovariant background field method to settle the
notation. In Section~\ref{sec_calculation} we calculate, to two loops,
the renormalized correlation function of two background gauge
superfields in supersymmetric gluodynamics. The contribution of matter
fields can also be computed using the techniques described here, but
the results will be presented elsewhere. In Section~\ref{sec_RGE} we
write the renormalization group equation and determine the first two
coefficients of the beta function. Section~\ref{sec_discussion} is
devoted to a discussion of the origin of the higher-order coefficients
in our calculation and in previous works. In
particular, we discuss the relation between our result and
the flow of the Wilsonian coupling constant. Finally, in the Appendix
we compute the two-point function of quantum gauge superfields at one
loop, and determine the renormalization group coefficient associated
to the gauge-fixing parameter.


\section{Differential renormalization}

\label{sec_DR}

Differential renormalization~\cite{Freedman:1991tk} is a method that
defines
renormalized correlation functions without intermediate
regulator or counterterms. This is achieved by writing
coor\-dinate-space expressions that are too singular as
derivatives of less singular ones. The derivatives are then
understood in the sense of distribution theory, \ie, they are
prescribed to act formally by parts on test functions, neglecting
divergent surface terms. Diagrams with subdivergences are
renormalized according to Bogoliubov's recursion formula.
This procedure leads to {\it bona fide\/} distributions that respect
the requirements of quantum field theory. Consider as a simple example
the singular function $(1/x^2)^2$. Its renormalized form is
simply expressed by
\be
\left[ \frac{1}{x^4} \right]_R = - \frac{1}{4} {\dal}
\frac{\ln x^2M^2}{x^2} + a_{UV} \delta(x) \label{basicidentity} \; ,
\end{equation}
where the D'Alembertian acts ``by parts''.
Note that the bare and the renormalized expressions, when understood
as functions, coincide for $x \not = 0$. However, only the
renormalized expression is a finite distribution on test functions
defined over the complete space. The arbitrary scale $M$,
with dimension of mass,
must be included for dimensional reasons and plays the r\^ole of a
renormalization group scale. Related to this is the fact that one can
always add a local term of the appropriate dimension, which reflects
the freedom in choosing a scheme. Note that the arbitrary constant
$a_{UV}$ can
be absorbed into a redefinition of $M$. Different renormalized
expressions can in principle have different renormalization scales
and/or different local terms. Our approach will be to write a single
(UV) renormalization scale and adjust the contact terms in such a way
that gauge invariance is preserved.

Analogously, IR divergent expressions can be made finite by
differential renormalization in momentum space. For instance,
\be
\left[ \frac{1}{p^4} \right]_{\tilde{R}} = - \frac{1}{4}{\dal}_p
\frac{\ln p^2/\bar{M}_{IR}^2}{p^2} + a_{IR} \delta(p)
\label{basicIRidentity} \; .
\end{equation}
We have defined for convenience $\bar{M}_{IR}=2M_{IR}/\gamma_E$,
where $\gamma_E$ is Euler's constant, and distinguished the IR scale
from the UV one. This is an explicit realization of the
so-called $\tilde{R}$ operation that subtracts IR
divergences. Again, diagrams with IR subdivergences are treated
according to a recursion formula~\cite{Chetyrkin:nn,Popov:1984xm}
analogous to the UV one.

Since UV and IR overall divergences are local in coordinate and
momentum space, respectively, the $R$ and $\tilde{R}$ operations
commute, and one can define an operation $R^*=\tilde{R}R$ to
renormalize both UV and IR
divergences~\cite{Chetyrkin:nn,Popov:1984xm}. The fact that the UV and
IR renormalizations decouple means that the UV and IR
renormalization scales should be independent. In DiffR this can be
achieved by a careful adjustment of the local terms involving both
scales\footnote{IR DiffR was investigated in~\cite{Avdeev:jp}
where it was concluded that the combination of UV and IR DiffR was
inconsistent, as the results depended on the order in which
integrations were performed. According to
\cite{Smirnov:1994km}, however, this corresponds to the
natural arbitrariness of the IR renormalization,
and this author
has actually proposed in~\cite{Smirnov:1996yi} a consistent version of
DiffR that deals with both
UV and IR divergences. Our approach here will be closer to the
original version of DiffR}. Let us implement this idea in an example
which will play
a central role in the calculation of Section~\ref{sec_calculation}.
Consider the IR singular expression
\be
\frac{\ln p^2/\bar{M}^2}{p^4} \; \label{singIR},
\end{equation}
that arises in IR divergent expressions after renormalization
of a UV subdivergence (with $a_{UV}$=0). The
consistent IR renormalization of~(\ref{singIR}) is given by
\be
\left[\frac{\ln p^2/\bar{M}^2}{p^4}\right]_{\tilde{R}} =
-\frac{1}{8} {\dal}_p \frac{-\ln^2 p^2/\bar{M}_{IR}^2 + 2 \ln
p^2/\bar{M}_{IR}^2 \, (1 + \ln p^2/\bar{M}^2)}{p^2} + (a_{IR} \ln
\frac{M_{IR}^2}{M^2} + b_{IR}) \delta(p)
\; . \label{identityUVIR}
\end{equation}
This expression differs from the usual one given
in~\cite{Freedman:1991tk} by scale-dependent local terms proportional to $\ln^2
M^2/M_{IR}^2$ (appart from the explicit local terms with coefficients
$a_{IR}$ and $b_{IR}$). It should be
used whenever the ``new'' scale is to be treated as independent from
the ``old'' one, for consistency of the loop expansion. The
scale-dependent local terms of~(\ref{identityUVIR}) are
fixed by the requirement that the IR renormalization commute with a
rescaling of $M$, that is to say,
\be
M \frac{\d}{\d M} \left[\frac{\ln
p^2/\bar{M}^2}{p^4}\right]_{\tilde{R}} =
\left[M \frac{\d}{\d M} \frac{\ln
p^2/\bar{M}^2}{p^4}\right]_{\tilde{R}} \; .
\end{equation}
Observe that the UV scale $M$ only appears in~(\ref{identityUVIR}) in
single logarithms. This is fine, for double logarithms of $M$ are
expected to appear only when the bare expression contains both
a UV subdivergence and a UV overall divergence. Observe also that a
rescaling of $M_{IR}$ in~(\ref{identityUVIR}) gives
a local term in $p$-space. This procedure can be
extended to more general situations, but the
identity~(\ref{identityUVIR}) is all we need for the calculation at
hand.

Let us finally deal with the scale-independent local terms. They must
be chosen such that the renormalized correlation functions respect the
fundamental symmetries of the theory. In our problem these are
supersymmetry and gauge invariance. Since the first is automatically
preserved in superspace, we only have to worry about the second. In
ordinary DiffR, we would have to study the Ward identities order by
order and adjust the local terms by hand so that they are
satisfied. For the calculation of the two-loop beta
function it is sufficient to impose the Ward identities at the
one-loop level, but this is complicated in the framework of covariant
supergraphs. Life gets much easier, however, when one uses the
so-called constrained differential renormalization
(CDR)~\cite{delAguila:1998nd}. This is a procedure that fixes the
arbitrary local
terms {\it a priori} in such a way that the Ward identities are
directly
fulfilled~\cite{delAguila:1998nd,delAguila:1997kw,
Perez-Victoria:1998fj}. Furthermore,
CDR respects supersymmetry in component field
calculations~\cite{delAguila:1997yd,delAguila:1997ma}. In superspace
calculations CDR is
particularly simple because, after performing the superalgebra, all
subdivergences are Lorentz scalars. According to the CDR
prescriptions, this means that the local terms in the renormalized
subdiagrams are universal, \ie, they are independent of the Green
function or diagram they appear in. Specifically, in our calculation
we shall take $a_{UV}=a_{IR}=0$, and keep $b_{IR}$ arbitrary
(but unique) till the very end.
When calculating beta functions, this is all we
need. Nevertheless, in order to calculate the complete two-point
function, we shall also fix by hand the local terms that
appear in superficially divergent tensor structures, so that
gauge invariance be preserved. (We need this straightforward adjustment
because CDR has not been developed beyond the one loop level.)


\section{Supercovariant background field method}

\label{sec_covariant}
In the background  field method, the total gauge superfield $V_T$ is
splitted into background $\bf B$
and quantum $V$ superfields according to
\begin{equation}
e^{V_T} = e^\bOmega e^{gV} e^{\bbOmega} ~~~~~~~~~~~;~~~~~~~~~~~
e^{\bf B}=  e^\bOmega e^{\bbOmega}
\end{equation}
We will follow closely the notation and conventions given in
\cite{Gates:nr}. Background covariant derivatives
can be defined as follows
\beqa
\bnabla_\alpha &\equiv& e^{-\bOmega}D_\alpha e^{\bOmega} = D_\alpha -i
\boldsymbol{\Gamma}_\alpha(\bOmega) \nonumber\\
\bbnabla_{\dot \alpha} &\equiv& e^{\bbOmega}\bar D_{\dot\alpha}
e^{-\bbOmega} =  \bar D_{\dot \alpha} -i
\boldsymbol{\bar \Gamma}_{\dot \alpha}(\bar \bOmega)
\eeqa
In the chiral representation for  covariant derivatives, the
``quantum-background" splitting amounts to
\beqa
\nabla_\alpha &=& e^{-V_T}D_\alpha e^{V_T} = e^{-\bbOmega} e^{-gV}{
\bnabla}_\alpha e^{gV}e^{\bbOmega} \nonumber \\
\bar\nabla_{\dot\alpha}  &=& \bar D_{\dot \alpha} = e^{-\bbOmega}
\bbnabla_{\dot \alpha}  e^{\bbOmega}~.
\eeqa
The classical action of pure N=1 SYM then adopts the
following form
\beqa
S_{YM} &=& \frac{1}{ g^2}\int d^4xd^2\th \,\Tr\, W^2 =
 \frac{1}{2g^2}\int d^4xd^2\th \,\Tr\left(
\frac{i}{2}\left[ \bar \nabla^{\dot\alpha},\left\{\bar
\nabla_{\dot\alpha},\nabla_\alpha \right\}\right]\right)^2 \nonumber
\\
&=&-\frac{1}{2g^2}\int d^4xd^2\th \,\Tr\left(\frac{1}{2}
\left[\bbnabla^{\dot\alpha}\left\{  \bbnabla_{\dot\alpha},e^{-gV}
\bnabla_\alpha e^{gV} \right\}\right]\right)^2
\label{ldkj}
\eeqa
The quantum-gauge fixing $ \bnabla^2 V = 0$ retains background covariance.
Usual averaging requires the introduction of Nielsen-Kallosh
ghosts:
\beqa
 & &\int {\cal D}f{\cal D}\bar f{\cal D}b{\cal D}\bar b~
\delta(\bnabla^2
V -f) ~\delta( \bbnabla^2  V -\bar f) ~\exp\left(  -
\frac{1}{\alpha }\int d^4 x d^4\th (f\bar f + b \bar b)\right)
\nonumber \\
& &=\int  {\cal D}b{\cal D}\bar b ~\exp \left(  -\frac{1}{ \alpha
 }  \int d^4 x d^4\th\, [\,\med V(\bnabla^2\bbnabla^2 +
\bbnabla^2 \bnabla^2)V +  b\bar b\,] \right)
\eeqa
Expanding $S_{YM}+S_{gf}$ in powers of the quantum field $V$ yields
\beqa
S &=& S_0+ S_2 + S_{int} \nonumber\\
&=& ~  \frac{ 1}{   g^2}\int d^4 x
d^2\th~\Tr {\bf W}^2  ~-~
\frac{1}{2}\int d^4 x d^2\th \,V[\hat{\dal} +\xi(\bnabla^2\bbnabla^2
+\bbnabla^2\bnabla^2)] V     \nonumber \\
 & & ~~~~~~- ~~  \int d^4xd^4\th \,\left(\bar c' c +
\bar c c' + (1+\xi) \bar b b    \right)~ +~ S_{int}(V,c,c')\label{llg}
\eeqa
where $\xi = \frac{1}{\alpha}-1$,
$\hat{\dal} = {\dal} -i\bGamma^{\al\dot\al}\d_{\al\dot\al}-
\frac{i}{2}(\d_{\al\dot\al}\bGamma^{\al\dot\al})
-\med\Gamma^{\al\dot\al}\Gamma_{\al\dot\al}
-i\bW^\alpha\bnabla_\alpha-i\bbW^{\dot\al}\bbnabla_{\dot\al}$ and the
dots stand for terms with higher powers of $V$ or $c,c'$.
All anticommuting superfields, $c,c'$ and $b$ interact with
the background field $\boldsymbol{\Gamma}$ through the constraint that
they be background covariantly chiral,
$
\bbnabla c = \bbnabla b = \bnabla \bar c = \bnabla \bar b = 0
$.
The Effective Action in the Background Field Gauge admits a
gauge invariant expansion in the form
\be
\Gamma[{\bf B}]= \int d^4x d^4 y \, d^2\th\left[ {\bf W}^\alpha(x){\bf
W}_\alpha(y)\right] G^{(2)}(y-x) + ... \label{ea}
\end{equation}
Our aim is to calculate the 2-point 1PI function $G^{(2)}$.
 For the perturbative computation of $\Gamma$, we expand the action in
powers of $V$ and distinguish the ``free'' part (in the presence of
the background), $S_0$, from the interacting part, $S_{int}$:
\beqa
\Gamma[{\bf B}] &=& S_0 +\Gamma^{1 loop}_\xi \nonumber\\
&& + \exp  S_{\rm
int}
\left[
\frac{\delta}{\delta J},
\frac{\delta}{\delta j},\frac{\delta}{\delta \bar j}  \right]
\exp \left[ \int d^4 x d^4 \th (\med J\hat{\dal}^{-1} J -\bar
j{\dal}_+^{-1} j   ) \right]_{J=j=\bar j=0}
\label{fullg}
\eeqa
where ${\dal}_+\Phi = \bar\nabla^2\nabla^2\Phi = [{\dal} - i {\bf W}^{\alpha} 
\bnabla_{\alpha} - \frac{i}{2} (\bnabla^{\alpha} {\bf W}_{\alpha})] \Phi$ is the D'Alembertian
acting on background-covariant chiral fields (similarly for   anti-chiral
fields ${\dal}_-\bar\Phi = \nabla^2\bar\nabla^2\bar\Phi =[ {\dal} - i \bbW^{\dot{\alpha}}
\bbnabla_{\dot{\alpha}} - \frac{i}{2}( \bbnabla^{\dot{\alpha}} \bbW_{\dot{\alpha}})]\bar\Phi$), 
and $J$, $j,\bar j$
  collectivelly  denote sources for vector  chiral
superfields  in (\ref{llg}).
$\Gamma^{1 loop}_\xi$ stands for the 1-loop contribution in the
gauge $\xi$.
We are interested in computing the two-point amplitude at two
loops in the Feynman gauge $\xi= 0$. However the gauge parameter
$\xi$ is renormalized and the RG equation will generically contain a term
$\gamma_\xi \d/\d\xi G^{(2)}|_{\xi = 0}$. Since $\gamma_\xi=
\d \xi/\d \log M = {\cal
O}(g^2) +...$ the linear dependence in $\xi$ of the 1-loop
Green's function will be needed.

\section{Calculation of the two-point function}

\label{sec_calculation}

\subsection{One Loop }
Formally, the exact one-loop contribution in an arbitrary Lorentz gauge,
$\xi$, is
\be
\Gamma^{1 loop}_\xi = -\med \Tr\ln \left(  \hat{\dal} +
\xi(\bnabla^2\bbnabla^2+ \bbnabla^2\bnabla^2) \right)
+  2\Tr\ln  (\bnabla^2
\bbnabla^2) +  \Tr\ln \left( (1 + \xi)   \bnabla^2
\bbnabla^2 \right) \; .
\end{equation}
The three terms represent a loop of quantum gauge superfields,
Faddeev-Popov ghosts and Nielsen-Kallosh ghosts,
respectively. Expanding to linear order in $\xi$
\be
\Gamma^{1 loop}_\xi = -\med \Tr\ln    \hat{\dal}
+  3\Tr\ln  (\bnabla^2
\bbnabla^2) + \frac{\xi}{2}\Tr \left[
\left(\frac{1}{{\dal}_-}-\frac{1}{\hat{\dal}}\right)
\bnabla^2
\bbnabla^2  + h.c.\right] +{\cal O}(\xi^2)  \label{adoiu} \; .
\end{equation}
From this expression, we are instructed to expand in powers of the
external background field $\bOmega$. The first term in (\ref{adoiu}) is
well known to start contributing from four-point  functions up
\cite{Gates:nr}.
The second piece, stemming from the ghosts, yields the standard
contribution to the 1-loop beta function in the
Feynman gauge.
 \bigskip
\begin{figure}[ht]
\centerline{\epsfbox{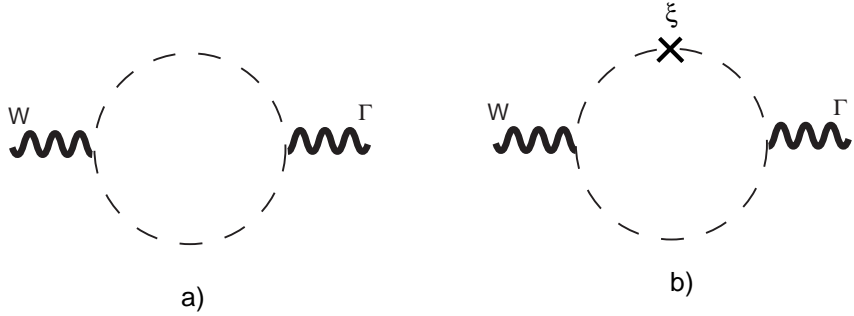}}
\caption{Diagrams contributing to $\Gamma_\xi^{1loop}$.}
\end{figure}
The diagram involved is shown in Figure 1-a) and yields
\cite{Song}
\beqa
&& \frac{3 C_A }{2}~\int d^4 x d^4 y d^4\theta \,
\bW^\alpha(x)\bGamma_\alpha(y) \Delta^2(x-y)  \nn
  & & ~~~~~ \stackrel{R}{\rightarrow} ~  -~\frac{3C_A }{ 4^3 \pi^2
}\int d^4 x d^4 y d^2\theta
\bW^\alpha(x)\bW_\alpha(y)   {\dal}\left[\Delta(x-y)\ln
(\xmy)^2 M^2\rule{0mm}{4mm} \right] \label{unlo}
\eeqa
after use of (\ref{basicidentity}) with $a_{UV}=0$. Here the free
propagator is  $\displaystyle
\Delta(x) \equiv \frac{1}{{\dal}}(x) = -\frac{1}{4\pi^2}
\frac{1}{x^2}$.
The last term in (\ref{adoiu}) involves corrections to the gauge
parameter. Standard $\bnabla$-algebra manipulations reduce it to
\beqa
-\frac{\xi C_A}{4}\int d^4xd^4y d^2\theta \, \bW^\alpha(x)\bW_{\alpha}(y)
{\dal}\left[\rule{0mm}{4mm} \Delta(x-y)\int d^4z
\Delta(z-y)\Delta(z-x)\right]
\eeqa
The integral that remains to be done corresponds to the insertion
diagram of figure 1b), and
diverges logarithmically for large $|z|$. This is the first instance
of an off-shell IR singularity, that we renormalize along the lines
explained in Section~\ref{sec_DR}:
\beqa
  \int d^4 z \; \frac{1}{(x-z)^2} \frac{1}{(z-y)^2}
&=&  \int d^4 p \;
\frac{1}{p^4} e^{-i p (\xmy)} \nonumber \\
&\stackrel{\tilde R}{\rightarrow}& - \frac{1}{4} \int d^4 p \; {\dal}_p
\frac{\ln \frac{p^2}{\bar{M}_{IR}^2}}{p^2} e^{-ip(x-y)} \nonumber \\
&=& -  \pi^2  \ln (\xmy)^2 M^2_{IR}  \;.\label{ert}
\eeqa
We have used the identity~(\ref{basicIRidentity}) with $a_{IR}=0$.
In the end we obtain the following contribution
to linear order in $\xi$:
\be
\frac{\xi C_A}{ 4^3 \pi^2 }  {\dal} [\Delta(x-y) \ln
(\xmy)^2 M_{IR}^2 ] \label{unloxi} \;,
\end{equation}
and the full one-loop contribution results in
\bea
\Gamma^{1loop}_\xi &=& - \frac{C_A}{4^3 \pi^2} \int d^4xd^4y d^2\theta \,
\bW^\alpha(x)\bW_{\alpha}(y)
{\dal} \left[ \rule{0mm}{4mm} 3\Delta(x-y) \ln ((x-y)^2 M^2) \right. \nn
&& ~~~~~~~~~~~~~ \left. \rule{0mm}{4mm}  \mbox{} -
\xi \Delta(x-y) \ln ((x-y)^2 M_{IR}^2)\right] +
{\cal O}(\xi^2) + \ldots  \label{oneloop}
\eea

\subsection{Two Loops}

We will compute the two-loop contribution in the Feynman gauge,
$\xi=0$. From
(\ref{fullg}), the higher-loop contributions come from
the expansion of a vacuum diagram with
propagators and vertices  made out of background covariant
derivatives (Figure 2). In particular there is a
single such diagram at two loops, as shown in \cite{Grisaru:ja}.
\bigskip
\begin{figure}[ht]
\centerline{\epsfbox{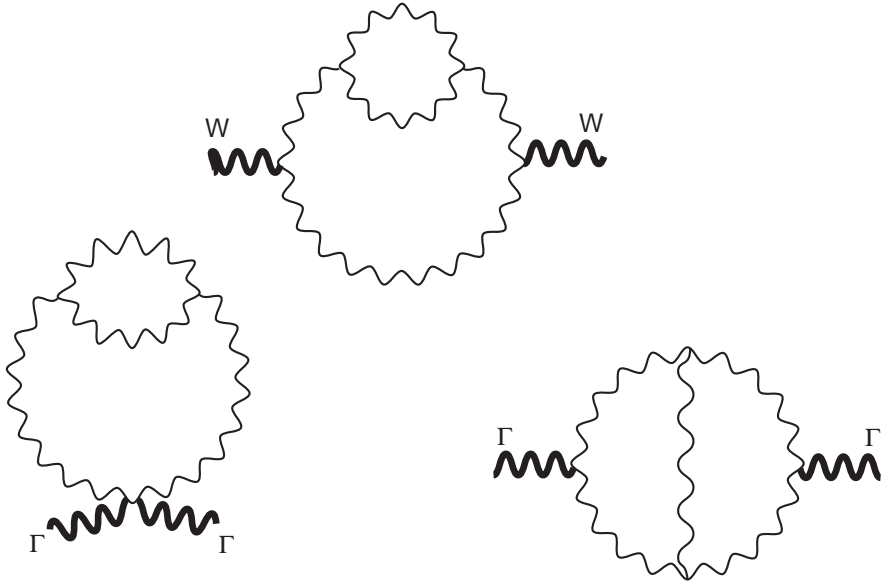}}
\caption{Diagrams contributing to $G^{(2)}$ at two loops.}
\end{figure}
\bigskip
After performing the $\bnabla$-algebra we find the following
nonvanishing contributions (up to a common factor $-3g^2
C^2_A/2$):
\beqa
D_1   &=&  i \int d^4 x d^4 y d^4 \th \; \bbW^{\dot{\al}}
(x, \th) \bW^{\al} (y, \th) ~\sigma^\mu_{\al\dot{\al}}~  [-  \Delta(\xmy)
\d_\mu I^0(x,y) ~] \; ,\nonumber \\
D_2  &=&  i \int d^4 x d^4 y d^4 \th \;
\bbW^{\dot{\al} }(x,\th) \bW^{\al} (y, \th)~\sigma^\mu_{\al\dot{\al}}~
\left[  2\Delta(\xmy) \d_\mu I^0(x,y) -  \d_\mu ( \Delta(\xmy) I^0(x,y) )
\right] \nonumber\\
D_3 &=&   \frac{1}{2} \int d^4 x d^4 y d^4 \th~
{\boldsymbol\Gamma}^\mu(x, \th)
{\boldsymbol\Gamma}^\nu (y, \th)  \left[-
 \d_\mu  \d_\nu  (\Delta I^0(x,y)) \right. \\ &&
~~~~~~~~~~~~~~~~~~~~~~~~~~~~~~\left. ~~~ + 4 \d_\mu
((\d_\nu\Delta)I^0(x,y)) - 4
((\d_\mu\d_\nu\Delta) I^0(x,y)) \right]   \; , \nonumber\\
D_{4} &=&  \med \int d^4 x d^4 y d^4 \th \;
{\boldsymbol\Gamma}^\mu ( x, \th) {\boldsymbol\Gamma}^\nu (y, \th) ~
\left( I_{\mu\nu}^1 (x,y) - I^{2 }_{\mu\nu} (x,y) \right) \;, \nonumber\\
D_{5} &=& \frac{1}{2} \int d^4 x d^4 y d^4 \th \;
{\boldsymbol\Gamma}^\mu ( x, \th) {\boldsymbol\Gamma}^\nu (y, \th)
~\delta_{\mu\nu}  ~({\dal}\Delta(\xmy))I^0(x,y)  \;.\nonumber
\eeqa
Here and in the  following, all derivatives act
on $x$ unless otherwise stated. In
$D_5$ we have replaced $\delta(x)$ by ${\dal}
\Delta(\xmy)$. The integrals $I^{i}$ are defined as follows:
\beqa
I^{0} (x,y) &=&   \int d^4 u d^4 v~ \Delta (x-u) \Delta (y-v) \Delta^2
(u-v)   ~~~~  \nonumber\\
I^{1}_{\mu\nu} (x,y) &=&\rule{0mm}{7mm}
 \int d^4 u d^4 v ~\Delta(u-v) \,  \Delta(x-v)
\d_{\mu}^x
\Delta(x-u)  \, \Delta(y-v) \d_{\nu}^y \Delta (y-u)        \\
I^{2 }_{\mu\nu} (x,y) &=& \int d^4 u d^4 v~ \Delta(u-v)\,  \Delta(x-u)
\d_{\mu}^x
\Delta(x-v) \,  \rule{0mm}{7mm}\Delta(y-v) \d_{\nu}^y \Delta (y-u)   .
\nonumber
\eeqa
As shown above, all expressions but $D_4$ are obtained from a single
integral $I^0$, which is both UV and IR logarithmically divergent.
On the other hand,  derivatives of $I^0(x)$  are just UV
divergent. The integrals $I^1$ and $I^2$, which appear in $D_4$,
are IR safe as well.

The strategy now is to look for a renormalized expression for the
sum $\sum_{i=1}^5 D_i$. To begin with we renormalize the integral
$I^0$. We must first cure the UV subdivergence, using the DiffR
identity (\ref{basicidentity}) with $a_{UV}=0$:
\br
I^0  &=& \frac{1}{(4 \pi^2)^4} \int  d^4 u d^4 v \;
\frac{1}{(x-u)^2} \frac{1}{(y-v)^2} \frac{1}{(u-v)^4} \;. \nn
&\stackrel{R}{\longrightarrow} & - \frac{1}{4 (4 \pi^2)^4} \int d^4 u
d^4 v \; \frac{1}{(x-u)^2}
\frac{1}{(y-v)^2} {\dal}^u \frac{\ln (u-v)^2 M^2}{(u-v)^2} \nn
&=& \frac{1}{4 (4 \pi^2)^3} \int d^4 v \frac{1}{(y-v)^2}
\frac{\ln (x-v)^2 M^2}{(x-v)^2} \nn
&=& - \frac{1}{4 (4 \pi^2)^3} \int d^4 p \; \frac{ \ln p^2/\bar{M}^2
}{p^4} e^{- i p(x-y)}\;.
\er
We are left with an IR divergent expression, which is renormalized
using the identity (\ref{identityUVIR}) (with $a_{IR}=0$). Integrating
by parts and Fourier transforming back into $x$ space we finally find
\be
I^0_R = \frac{1}{32 (4 \pi^2)^2} \left[ \ln^2 (x-y)^2 M^2_{IR} +
2 \ln (x-y)^2 M^2_{IR} \left( 1 - \ln (x-y)^2 M^2 \right)
+b_{IR}\right]  \label{I0}
\end{equation}
Using this result we readily obtain, with $\kappa=1/(4 \pi)^4$,
\beqa
D_1+ D_2 &=& - i \int d^4 x d^4 y d^4 \th \; \bW^{ \al }
(x, \th) \bbW^{\dot\al} (y, \th)\times
\nonumber\\
&& ~~~~ \sigma^\mu_{\al\dot{\al}}\d_\mu\left(
-\kappa \Delta(x\!-\!y)\ln(x\!-\!y)^2 M^2  +
\Delta(x\!-\!y)  I^0_R(x\!-\!y) \right) \;, \label{preD1D2}
\eeqa
which, through the use of Bianchi identities $\bnabla^\alpha\bW_\alpha =
\bbnabla^{\dot\alpha}\bbW_{\dot\alpha}$, can be
written as an F term:
\beqa
D_1+ D_2 &=& - \int d^4 x d^4 y d^2 \th \; \bW^{ \al }
(x, \th) \bW_{\al} (y, \th) \times\nonumber\\
&&~~~~~~{\dal} \left(
 \kappa\Delta(\xmy)\ln(x\!-\!y)^2 M^2 -
\Delta(\xmy)  I^0_R(\xmy) \right) \;.
\eeqa
To calculate $D_3+D_5$ it is convenient to decompose the last term
of $D_3$ (which has an overall divergence) into trace and traceless
parts:
\be
4\left[(\d_\mu\d_\nu\Delta)I^0\right]_R =
\left[((4\d_\mu\d_\nu-\delta_{\mu\nu} {\dal})\Delta)I^0\right]_R +
\delta_{\mu\nu} \left[({\dal} \Delta)I^0\right]_R + c \,\delta_{\mu\nu}
\delta(\xmy) \;.
\end{equation}
According to CDR, we have included a local term that can appear in the
trace-traceless decomposition at the renormalized
level~\cite{delAguila:1998nd}. We shall fix its
coefficient
later on requiring gauge invariance. Up to this term, the trace
cancels the complete diagram $D_5$. The rest of $D_3$
is  overall UV finite. Renormalizing the
subdivergences we find, after some algebra,
\beqa
D_3+D_5 &=& \int d^4x d^4y d^4\th~ \bGamma^\mu(x,\th) \bGamma^\nu(y,\th)
 \nonumber\\
&& \times \mbox{}  \left\{\rule{0mm}{7mm}
\right.(\d_\mu\d_\nu -\delta_{\mu\nu}{\dal}) \left(-\frac{1}{2}
\Delta(\xmy) I^0_{\cal R}(\xmy)  + \kappa\,\Delta(\xmy)\left[ \ln
 (\xmy)^2{M}^2+\med\right] \right) \nonumber\\
&& ~~~ +~\delta_{\mu\nu} \, \left[ \frac{\kappa}{4}\, {\dal}\left(
\Delta(\xmy)\ln (\xmy)^2{M}^2 \right)
 + e  \,\delta(\xmy)
\right]
\left.\rule{0mm}{7mm}
\right\} . \label{D3D5}
\eeqa
With $e= \frac{3}{8} \kappa + \frac{c}{2}$. The non-trasverse pieces
in the last line must be cancelled by
$D_4$. Let us consider this contribution next.
Again, it is convenient to split it into traceless and trace parts:
\be
I^1_{R\,\mu\nu}- I^2_{R\,\mu\nu}=
(I^1_{\mu\nu}-\frac{\delta_{\mu\nu}}{4}  I^1) -
(I^2_{\mu\nu}-\frac{\delta_{\mu\nu}}{4}  I^2) +
\frac{\delta_{\mu\nu}}{4} (I^1_{R\,\rho\rho}-I^2_{R\,\rho\rho}) +
c^\prime \delta_{\mu\nu} \delta(\xmy)
\, .
\end{equation}
We have included again an arbitrary local term in the trace-traceless
decomposition.
The trace can be computed using Gegenbauer
polynomials~\cite{Haagensen:1991vd,Song}. (Alternatively, its scale
dependent part
can be easily obtained with ``systematic''
DiffR~\cite{Latorre:1993xh}.)
The traceless part is UV and IR finite. Therefore, it is fixed by
dimensionality and by the traceless condition to be of the form
$a (\d_\mu\d_\nu-\frac{\delta_{\mu\nu}}{4}{\dal})\frac{1}{(\xmy)^2}$.
The coefficient $a$ may be determined from a rather cumbersome
calculation with Feynman parameters. Adding
trace and traceless parts we find
\bea
I^1_{R\,\mu\nu} - I^2_{R\,\mu\nu} &=& a (\d_\mu\d_\nu -
\delta_{\mu\nu}{\dal})  \Delta(\xmy) \\
&&  - \,\delta_{\mu\nu}{\dal}
\left[ \Delta(\xmy) \left( \frac{\kappa}{4} \ln \,
(\xmy)^2
\!   {M}^2   - (\frac{3}{4} \zeta(3) \kappa + c' + \frac{3}{4}a)  \right)
\rule{0mm}{5mm}\right] \nonumber \; .
\eea
Adding this to~(\ref{D3D5}) we obtain
\bea
D_3+D_4+D_5 &=& -\frac{1}{2} \, \int d^4x d^4y d^4\th~ \bGamma^\mu(x,\th)
\bGamma^\nu(y,\th) \nonumber \\
&& ~~~ \times ~ (\d_\mu\d_\nu
-\delta_{\mu\nu}{\dal})
\left[\rule{0mm}{5mm}
\Delta(\xmy) \left(I^0_{\cal R}(\xmy) -2 \, \kappa  \ln
(x\!-\!y)^2{M}^2 - a \right) \right] \nonumber \\
&& + \; \frac{h}{2}
\int d^4x d^4y d^4\th~ \bGamma^\mu(x,\th)
\bGamma^\nu(y,\th) \delta_{\mu\nu}
 {\dal} \Delta \label{preD3D4D5}  \; ,
\eea
with $h = \frac{3}{4} \kappa (1 + \zeta (3)) + \frac{3}{4}a + c +
c'$. The $M$- and $M_{IR}$-dependent parts are automatically
transverse, and the non transverse contribution of the scale
independent local part can be set to zero (so that the sum is gauge
invariant) adjusting $c$ and $c'$ appropiately.
Thanks to its transversality, we can then rewrite
this expression in a form proportional to the classical action:
\bea
D_3+D_4+D_5 &=&
-\frac{3}{4} \,
\int d^4x d^4y d^2\th~ \bW^\al(x,\th) \bW_\al(y,\th) \\
&&  ~~~\times~
{\dal}\left[\Delta(\xmy) \left( \rule{0mm}{4mm}
I^0_{\cal R}(\xmy)) -   2  \kappa\ln
(\xmy)^2{M}^2 - a \right)\right]. \nonumber
\eea
Summing up all diagrams and using~(\ref{I0}), we finally obtain for
the full 2-loop
contribution in the Feynman gauge $(\xi = 0)$
\beqa
\frac{- 3g^2 C_A^2}{2}\,\sum_{i=1}^5D_i &=&
 - \frac{3 g^2 C_A^2}{4^6 \pi^4}\int d^4x d^4y d^2\th~ \bW^\al(x,\th)
\bW_\al(y,\th)~  {\dal}\,\left[\rule{0mm}{4mm}\Delta(x-y)
\right.  \label{trmcumci}  \\
 && \hspace{-2cm} \times~ \left. \left(\ln^2(x\!-\!y)^2M_{IR}^2 + 2 \ln
(x\!-\!y)^2M_{IR}^2 (1-\ln (x\!-\!y)^2 M^2) + 4 \ln
(x\!-\!y)^2M^2\rule{0mm}{4mm}\right) \right] \nonumber
\eeqa
A possible local and scale-independent contribution has been
cancelled by an adequate choice of $b_{IR}$ in (\ref{I0}).
We see that only single logarithms of $M$ appear in the final
result, as required by renormalization group invariance (see
Section~\ref{sec_RGE}). This fact can also be understood in the following
way: For a double $M$
logarithm we must have both UV subdivergences and overall UV
divergences. By power counting, only the $\bGamma \bGamma$ terms
may
contain overall UV divergences. Furthermore, the traceless parts
multiplying $\bGamma \bGamma$ are finite  and can have only single
logarithms of $M$, arising from the subdivergences. But gauge
invariance, {\em i.e.} transversality, forces the trace part to have the same logarithm
structure, so double logarithms of $M$ are also forbidden in the complete result. The
same argument shows that were the theory scale independent (\ie,
finite) to $n$ loops, the background two-point function would also be
scale independent to $n+1$ loops\footnote{The corresponding properties for
the $1/\eps$ poles in DRed were found in~\cite{Grisaru:1985tc}.}. This
line
of reasoning can be pushed even further if we distinguish in
our calculation the two-loop UV scale, $M^\prime$, from the one-loop
scale, $M$. Requiring transversality on $M$ and $M^\prime$
independently we see that $M^\prime$ must cancel in the final
result. (This means that the coefficient $c$ contains the
expression $\ln {M^\prime}^2/M^2$.) So, it is only the one-loop scale
that appears in the renormalized two-loop function. The significance
of this observation is discussed in Section~\ref{sec_discussion}.

\indent

\section{Renormalization group equation}
\label{sec_RGE}
After adding up the partial results~(\ref{oneloop})
and~(\ref{trmcumci}), the final renormalized expression for the
background two-point function reads
\bea
G^{(2)}(x) &=& \frac{1}{2g^2}\delta(x) \nonumber\\
&& \rule{0mm}{8mm} + \frac{3C_A}{4^2 (4\pi^2)^2 }
{\dal} \frac{\ln(x^2 M^2)}{x^2}
- \frac{\xi  C_A}{ 4^2 (4 \pi^2)^2 }
{\dal} \frac{\ln(x^2M_{IR}^2)}{x^2} +
{\cal O}(\xi^2) \nonumber\\
&&\rule{0mm}{8mm}
+ \frac{3 g^2 C_A^2}{4^4 (\cpc)^3 } \left\{ \rule{0mm}{6mm} {\dal}
\frac{
 \ln^2 (x^2M_{IR}^2) + 2\ln (x^2 M_{IR}^2) (1-\ln (x^2M^2))  + 4
 \ln (x^2 M^2)}{x^2} \right.
\nonumber \\
&& \rule{0mm}{8mm} + {\cal O}(\xi) \left. \rule{0mm}{6mm} \right\} +
{\cal O}(g^4) \; .
\eea
Due to background gauge invariance, $\bW$
undergoes no wave function renormalization. Thus the renormalization
group equation for this Green function reads
\be
\left.\left( M \frac{\d}{\d M} + \beta (g) \frac{\d}{\d g}
+\gamma_{\xi}(g)
\frac{\d}{\d \xi}
\right) G^{(2)} (x)\right\vert_{\xi =0} = 0 \;. \label{CS}
\end{equation}
Note that we can only go into the Feynman gauge after evaluating the
derivative with respect to $\xi$.
We have not included a term $\gamma_{IR} M_{IR} \d /\d M_{IR}$ because
in deriving the renormalization formula (\ref{identityUVIR}) we
required that the UV and IR scales were independent from each
other, and hence $\gamma_{IR} = M/M_{IR} \, \d M_{IR}/\d M =0$. In
fact, $M_{IR}$ parametrizes nonlocal contributions, which are
not object of UV renormalization.

We solve the renormalization group equation perturbatively to order
$g^2$ (two loops).
The first coefficient in the expansion of $\gamma_\xi(g)$ can be read
off from the 1-loop vacuum
polarization for the gauge superfield field $V$, which is
calculated in the Appendix with the result $\gamma_\xi(g) = -
\frac{3C_A}{4 (4\pi^2)} g^2 + O(g^4)$. With this input, all the nonlocal
and scale dependent pieces cancel out in (\ref{CS}). This is a check
of the consistency of our renormalization
procedure. In particular, note that the IR scale generated in the
gauge parameter at two loops is cancelled by corrections to the
gauge parameter at one loop. The remaining local parts of the
renormalization group equation~(\ref{CS}) uniquely fix the first two
coefficients of the beta function:
\footnote{ A word on normalization. The coupling constant $g$ as given in
(\ref{ldkj}) is $\sqrt 2$ times larger than the standard
Yang-Mills coupling, $g_{YM}$ (see p.\ 55\, in \cite{Gates:nr}). In
terms of the latter,
$
\beta (g_{YM}) =
-\frac{3}{2}\frac{ C_A }{8 \pi^2}g_{YM}^2 - \frac{3}{2}(\frac{C _A }{
8\pi^2 })^2 g_{YM}^5
+ \ldots
$
which matches the expansion of expression (\ref{NSVZ}).
}
\begin{equation}
\beta_(g) =  -\frac{3}{4} \frac{ C_A}{8\pi^2}g^3 - \frac{3}{8}
\left[\frac{ C_A}{8\pi^2}\right]^2 g^5 + {\cal O}(g^7)
\end{equation}

\section{Discussion}

\label{sec_discussion}

Using DiffR, we have calculated the complete
renormalized correlation function of two background gauge superfields
in pure N=1 SYM. This calculation illustrates the
power and simplicity of this method in applications to
supersymmetric gauge theories. In particular, we have seen that DiffR
can be employed to subtract IR divergencies as well, and that the
corresponding IR scale can be clearly distinguished from the UV one.
From the dependence of the two-point function on the
renormalization scale we have derived the first and second
coefficients of the beta function, $b_0$ and $b_1$. Furthermore, we
have presented a new argument showing that the $n+1$-loop coefficient
vanishes for any supersymmetric theory which is
finite to $n$ loops. This important property had been proven before
using DRed~\cite{Grisaru:1985tc}.

It is interesting to have a closer look at the way in which the
two-loop coefficient of the beta function, $b_1$, is generated in our
calculation.
\begin{enumerate}
\item UV one-loop subdivergencies are subtracted. This entails
a one-loop wave-function renormalization of the quantum gauge
superfield. The corresponding renormalized subdiagrams depend on
the one-loop renormalization scale $M$ in a local way. 
\item The overall UV
divergencies are subtracted and a new renormalization scale
$M^\prime$ is introduced. However, the combination of supersymmetry,
gauge invariance and power counting implies that $M^\prime$ cancels
out in the complete renormalized two-point function. On the other
hand, there remains a nonlocal dependence on $M$ (see
Eqs. (\ref{preD1D2}) and~(\ref{preD3D4D5})).\footnote{Note that the
expression ``dependence on $M$'' refers to the derivative with respect
to $M$ and not to the term in which $M$ appears, which is always
nonlocal.}
\item After integration over half the supercoordinates,
the dependence on $M$ becomes local.
\item Finally, this local scale dependence is compensated by $b_1$ in
the renormalization
group equation. The off-shell IR divergencies only play a passive r\^ole,
as they exactly cancel in the renormalization group equation.
\end{enumerate}
Summarizing, the scale associated to the one-loop renormalization
of the quantum superfield is the one that gives rise to the two-loop
coefficient of the beta function! This is somewhat surprising because
na\"{\i}vely one would think that the two-loop coefficient should have
its origin in $M^\prime$, which is the scale associated to two-loop
superficial divergencies. Should this be the case, the beta function
would be purely one loop (remember that no overall scale can appear at
any order $n>1$). However, we have seen explicitly in a two-loop
calculation that the subdivergences play a nontrivial role. More
generally, we expect that subdivergencies are responsible for all
higher-order coefficients of the beta function. This agrees with the
NSVZ form of the beta function. The fact that $M^\prime$ disappears in our
method is directly related to the observation in
\cite{Grisaru:1985tc} that
in invariant four-dimensional regularization methods there are no
divergencies after all subdivergencies have been subtracted. As we
have seen, this does not imply $b_1=0$. Therefore, it seems that
the na\"{\i}ve perturbative derivation of the beta function from
renormalization constants
needs some modification in this case. This modification is surely
related to the presence of the anomaly discussed
in~\cite{Shifman:1986zi,Arkani-Hamed:1997mj,
Kraus:2001tg,Kraus:2001id}. From
this point of view, the fact that the standard derivation from
renormalization constants works in DRed seems related to the fact
that there are no rescaling anomalies in this
method~\cite{Gates:nr}.

As a matter of fact, the mechanism we have just described agrees with
previous calculations in which the corrections to the one-loop
result arise from a one-loop
anomaly~\cite{Shifman:1986zi,Arkani-Hamed:1997mj,
Kraus:2001tg,Kraus:2001id}.
This anomaly manifests itself in different ways: as a
nonzero expectation value for the operator $WW$~\cite{Shifman:1986zi};
as the noninvariance of the measure under the rescaling of the gauge
field~\cite{Arkani-Hamed:1997mj}; or
as the quantum breaking of either supersymmetry or holomorphy in the
framework of local coupling~\cite{Kraus:2001tg,Kraus:2001id}.
In our description, the anomaly is to be associated with the external
loop, and is responsible for the promotion of the $M$ dependence into
a non-vanishing nonlocal structure that eventually generates
$b_1$. This is completely analogous to the explicit calculations
in SQED performed in~\cite{Shifman:1986zi} (except for the fact that
we subtract the subdivergencies). Note that even though $M'$ cancels
out, the presence of a $\infty - \infty$ UV behaviour is crucial for
the anomaly to exist. On the other hand, as emphasized
in~\cite{Shifman:1986zi}, this nonlocal structure is nonanalytic at
vanishing external momentum. Such on-shell IR divergence, which
cancels after integration over the $\bar{\theta}$ coordinates, is a
manifestation of the ``IR side'' of the
anomaly~\cite{Shifman:nw} and should not be confused with the off-shell
IR divergences that we have renormalized in our calculation. More
generally, IR effects are known to be responsible for quantum
corrections to F terms in the 1PI effective action~\cite{West:qt}.
The on-shell IR divergence arises from the region of virtual momenta
of order the momentum of the external field. If this region is
excluded, one finds $b_1=0$. Our subtraction of IR divergencies,
on the other hand, is local in momentum space and
does not modify the analiticity properties of the two-point
function. The same applies to
the IR subtractions in~\cite{Grisaru:1985tc}. Summarizing, {\em the
two-loop coefficient of the beta function arises from a one-loop UV
scale which survives at two loops only when IR effects are included}.

For completeness, we discuss in the rest of this section the
connection between the Gell-Mann-Low beta function we have computed
(1PI beta function) and the flow of the coupling in the Wilsonian
action (Wilsonian beta function). The Wilsonian effective action can
be understood as the generating functional of Green functions with an
IR cutoff, for external momenta smaller that this
cutoff~\cite{Keller:1990ej}. Therefore, at
least in perturbation theory, the (``holomorphic'') Wilsonian
coupling obeys a one-loop renormalization group flow.
(For pure Super Yang-Mills this result also holds when nonperturbative
effects are included.) The higher-order contributions to the physical
beta function appear when calculating expectation values of the operators
in the Wilsonian action~\cite{Shifman:1986zi}. On the other hand, in
~\cite{Arkani-Hamed:1997mj}
the ``canonical'' Wilsonian coupling was shown  to obey instead an NSVZ
flow. According to this work, in a Wilsonian setup this comes
about because a rescaling of the gauge
superfields at each scale is needed in order to keep kinetic
terms  canonically normalized. This rescaling induces an anomalous
Jacobian, and it is this anomaly that induces the corrections to the
one-loop result~\cite{Arkani-Hamed:1997mj}. In this reference, the
anomaly is calculated {\it \`a la\/} Fujikawa. A basic element of the
calculation is the introduction of a UV cutoff and one might believe
that the anomaly (and thus the running of the canonical
coupling) depends only on the UV properties of the theory. However, a
closer look shows that the low-energy degrees of freedom play a
fundamental r\^ole~\cite{Shifman:1999kf,Shifman:nw}. In fact, the
IR degrees of freedom must be included in the derivation of the
anomaly if low-energy physics is to remain unchanged under the field
rescalings. In this sense, taking the rescaling anomaly into account
is equivalent to calculating the expectation value of the Wilsonian
action~\cite{Shifman:1986zi}.

It was also argued in~\cite{Arkani-Hamed:1997mj}
(see~\cite{Arkani-Hamed:1997ut} as well)
that the canonical Wilsonian coupling is closely related to the 1PI
running coupling. (This implies that the expectation value of the
canonical Wilsonian action is basically trivial.) This relation has
been made more precise, in a general context,
in~\cite{Bonini:1996bk}. There it is shown that,
when all kinetic terms are canonically normalized, the
Wilsonian coupling becomes independent of the renormalization
scale for large cutoff. Then one can derive the equation (see
also~\cite{Collins})
\be
\beta(g) = \left[\frac{\Lambda \d /\d
\Lambda \, g_{cW}(g,\Lambda/M)}{\d /\d g \, g_{cW}(g,\Lambda/M)
}\right]_{\Lambda\rightarrow
\infty} \; . \label{betaphys}
\end{equation}
Here $g$ is the physical coupling, $g_{cW}$ is the canonical Wilsonian
coupling, $\Lambda$ is the flowing cutoff in the Wilsonian action and
$M$ is the renormalization scale, introduced by low-energy
normalization conditions. Using this equation, we show now that
(at least for large $\Lambda$) the 1PI and the canonical
Wilsonian beta functions agree to two loops in perturbation theory, as
functions of $g$ and $g_{cW}$, respectively.
The flow of $g_{cW}$ is of the generic form
\be
\beta_{cW}(g_{cW})\equiv \Lambda \d g_{cW}/\d \Lambda = b_0 g_{cW}^3 +
b_1 g_{cW}^5 + b_2 g_{cW}^7 + {\cal O}(g_{cW}^9) \; , \label{betacW}
\end{equation}
with constant coefficients $b_n$. On the other hand, $g_{cW}$ can be
perturbatively expanded in powers of $g$:
\be
g_{cW} = g \left(1 + C_0(\Lambda/M) g^2 + C_1(\Lambda/M) g^4 +
{\cal O}(g^6)\right) \; , \label{gcW}
\end{equation}
where we have taken $g_{cW}=g$ at tree level.
Inserting (\ref{gcW}) into (\ref{betacW}) we see that
\bea
C_0(\Lambda/M) &=& b_0 \ln \frac{\Lambda}{M} + c_0 \;, \nn
C_1(\Lambda/M) &=& \frac{3}{2} b_0^2 \ln^2 \frac{\Lambda}{M} + (2b_0
c_0 + b_1) \ln \frac{\Lambda}{M} + c_1 \, , \label{constants}
\eea
where $c_n$ are scheme dependent constants. Using (\ref{betacW}) and
(\ref{gcW}) in Eq.~(\ref{betaphys}) we find
\be
\beta(g) = \left[b_0 g^3 + b_1 g^5 + (b_2 + 2 b_1 C_0 + 3 b_0 C_0^2 -
2 b_0 C_1) g^7 + {\cal O}(g^9)\right]_{\Lambda\rightarrow \infty} \; ,
\end{equation}
and using also (\ref{constants}) we see that the beta function is
finite:
\be
\beta(g) = b_0 g^3 + b_1 g^5 + (b_0 (3c_0^2 - c_1) + b_1 c_0 + b_2)
g^7 + {\cal O}(g^9) \; . \label{finalbeta}
\end{equation}
Hence, the first two coefficients of the 1PI beta function coincide, in
any (mass-independent) scheme, with the first two coefficients of the
canonical Wilsonian beta function. The other coefficients are scheme
dependent, as expected. This scheme dependence has been studied for
general N=1 theories in~\cite{Jack:1996cn}.

\vspace{1cm}


\section*{Acknowledgements}

It is a pleasure to thank D.Z. Freedman, M.T. Grisaru, J.I. Latorre
and D. Zanon for useful discussions.
The work of J.M. was partially supported by DGCIYT under
contract PB96-0960. The work of M.P.V. was supported in part by the
European Program HPRN-CT-2000-00149 (Physics at Colliders).

\vspace*{.5cm}

\appendix

\section{Calculation of $\gamma_{\xi}$}

The classical action in a generic Lorentz gauge reads
\br
S &=& - \frac{1}{2 g^2} \int d^8 z \; (e^{- g V} D^{\a} e^{g V})
\bar{D}^2 ( e^{-g V} D_{\a} e^{g V}) - (\xi +1) \int d^8 z \; (D^2 V)
(\bar{D}^2 V) \;.
\er
To second order in the quantum field $V$ it reduces to
\br
S^{(2)} &=& \frac{1}{2} \int d^8 z \; V D^{\a} \bar{D}^2 D_{\a} V -
\frac{(\xi +1)}{2} \int d^8 z \; V ( D^2 \bar{D}^2 + \bar{D}^2 D^2 ) V
\nonumber \\
&=& - \frac{1}{2} \int d^8 z \; V {\dal} \Pi_{1/2} V - \frac{(\xi
+1)}{2} d^8 z V {\dal} \Pi_{0} V \;,
\er
were we have defined the projectors
\br
\Pi_{1/2} &=& - \frac{D^{\a} \bar{D}^2 D_{\a}}{{\dal}} \\
\Pi_{0} &=& \frac{D^2 \bar{D}^2 + \bar{D}^2 D^2}{{\dal}} \; .
\er
The one-loop correction to this quadratic action can be easily
computed. The result in the Feynman gauge is
\br
\Gamma &=& - \frac{ 3 C_{A} g^2}{4 (4 \pi^2)^2} \int d^8 z_1 d^8 z_2
\; V(z_1) D^{\a} \bar{D}^2 D_{\a} V(z_2) \delta_{12}
\frac{1}{(x_1-x_2)^4} \nonumber \\
& \stackrel{R}{\rightarrow}& - \frac{3 C_A g^2}{16 (4 \pi^2)} \int d^8
z_1 d^8 z_2 \; V(z_1) D^{\a} \bar{D}^2 D_{\a} V(z_2) \delta_{12} {\dal}
\left[ \Delta (x_{12}) \ln x_{12}^2 M^2 \right] \;.
\er
Therefore, the one loop effective action quadratic in the
quantum gauge field is
\br
\Gamma &=& \int d^8 z_1 d^8 z_2 \; V(z_1) \Gamma^{(2)}(z_1,z_2) V(z_2) \nn
&=& \int d^8 z_1 d^8 z_2 \; V(z_1) \left[ - \frac{1}{2} \delta^{(8)}
(z_{12}) {\dal} \Pi_{1/2} - \frac{(\xi+1)}{2} \delta^{(8)} (z_{12}) {\dal}
\Pi_{0} - \nonumber \right. \\
&+& \left.\frac{3 C_A g^2}{16 (4 \pi^2)} \left( {\dal} \left[
\Delta(x_{12}^2) \ln x_{12}^2 M^2 \right] {\dal} \Pi_{1/2} + {\cal
O}(\xi) \right) \right] V(z_2) \;.
\er
The renormalization group equation for the correlation function of two
quantum gauge fields has the form:
\br
\left. \left(M \frac{ \partial}{\partial M} + \b
\frac{\partial}{\partial g}
+ \gamma_{\xi} \frac{\partial}{\partial \xi} - 2 \gamma_{V}  \right)
\Gamma^{(2)} \right|_{\xi=0} =0 \;.
\er
To order $g^2$ (one loop) it is solved for
\br
\gamma_{V} &=& - \frac{3 C_A}{8 (4 \pi^2)} g^2 + \ldots \\
\gamma_{\xi} &=& - \frac{3 C_A}{4 (4 \pi^2)} g^2 + \ldots
\er


\end{document}